\documentclass[a4paper,12pt,twoside]{article}
\title{\textbf{Standard Bases for Linear Codes over Prime Fields}}
\author{Jean Jacques Ferdinand RANDRIAMIARAMPANAHY$^{(1)}$\\e-mail : randriamiferdinand@gmail.com\\Harinaivo ANDRIATAHINY$^{(2)}$\\e-mail : hariandriatahiny@gmail.com\\Toussaint Joseph RABEHERIMANANA$^{(3)}$\\e-mail : rabeherimanana.toussaint@yahoo.fr\\\\$^{1,2,3}$Mention : Mathematics and Computer Science,\\Domain : Sciences and Technologies,\\University of Antananarivo, Madagascar}
\usepackage{geometry}
\usepackage{amsmath,amssymb}
\usepackage{amscd,amsfonts}
\usepackage{amsthm}
\usepackage[english]{babel}
\usepackage[T1]{fontenc}
\usepackage[ansinew]{inputenc}
\swapnumbers\theoremstyle{plain}
\swapnumbers
\theoremstyle{plain}

\usepackage[pdfborder={0 0 0}]{hyperref}
\usepackage{hyperref}

\newtheorem{thm}{Theorem}[section]
\newtheorem{prop}[thm]{Proposition}
\newtheorem{lem}[thm]{Lemma}

\DeclareMathOperator{\red}{Red}
\DeclareMathOperator{\lt}{\mathrm{lt}}
\DeclareMathOperator{\spoly}{\mathrm{spoly}}
\DeclareMathOperator{\ecart}{\mathrm{ecart}}

\DeclareMathOperator{\lc}{\mathrm{lc}}
\DeclareMathOperator{\lm}{\mathrm{lm}}
\DeclareMathOperator{\loc }{Loc}
\DeclareMathOperator{\enn}{\mathbb{N}}

\theoremstyle{definition}

\newtheorem{ex}[thm]{Example}

\DeclareMathOperator{\card}{card}
\DeclareMathOperator{\lcm }{lcm}
\DeclareMathOperator{\supp }{supp}

\newcommand{\pr}{\noindent\textit{\textbf{Proof. }}}

\begin{document}

\maketitle

\begin{abstract}
It is known that a linear code can be represented by a binomial ideal. In this paper, we give standard bases for the ideals in a localization of the multivariate polynomial ring in the case of the linear codes over prime fields.
\end{abstract}

\section{Introduction}
 Coding theory is the mathematical basis for data  transmission through noisy communication channels. It contains  two main parts. The first part is to encode the message to reduce its sensitivity to noise during transmission. The second part is to
 decode the received message by detecting and correcting the errors.
  
 Bruno Buchberger introduced the theory of Groebner bases for polynomial ideals in $1965$. The Groebner bases theory can be used to solve some problems concerning the ideals by developing computations in multivariate polynomial rings. In $1964$, Hironaka \cite{hironaka} introduced the  analogues of Groebner bases called standard bases for ideals in the localization of the polynomial ring at the origin. In \cite{clo2}, standard bases for ideals generated by polynomials in local rings can be determined by using the same method as Groebner bases.
 
  Connection between linear codes and ideals in polynomial rings
 was presented in \cite{borg4}. And it was proved that a Groebner basis of the ideal associated to a binary linear code  can be used for determining the minimum distance . A generalization to linear codes over prime fields can be found in \cite{zi1,zi2}. 
In \cite{zi1}, it has been proved that a linear code can be described by a binomial ideal, and a Groebner basis with respect to lexicographic order for the binomial ideal is determined .
 
 The aim of this paper is to present the standard basis of the ideal of a linear code over a prime field in the local ring of rational fonctions that are regular at a point of the affine variety associated to the ideal. The  idea is to generalize the method developed by N. D\"uck and K. H. Zimmermann in \cite{f2}.
 
 This paper is organized as follows. The second section presents a background for Groebner bases. The third section contains the division algorithm in a local ring. In section $4$, the notion of linear codes and their connections with binomial ideals are presented. The main result is  contained in section $5$.

\section{Preliminaries}
Throughout this paper, $n$ denotes a positive integer, $\mathbb{K}$ a commutative field and $\mathbb{K}[X]:=\mathbb{K}[X_1,\dots,X_n]$ the polynomial ring in $n$ variables over $\mathbb{K}$. We denote by $0$ the zero element of $\enn^n$ where $\mathbb{N}$ is the set of non negative integers. \\ 
 A monomial in $\mathbb{K}[X]$ is an algebraic expression of the form $ X_1^{\alpha_1}\cdots X_n^{\alpha_n}$ which is denoted by $ X^{\alpha}$ where $\alpha=(\alpha_1,\ldots,\alpha_n)\in\enn^n$. 
  The monomial $X^{\alpha}= X_1^{\alpha_1}\cdots X_n^{\alpha_n}$ can be identified with the n-tuple of exponents $\alpha=(\alpha_1,\ldots,\alpha_n)\in\enn^n$and vice versa, thus there exists a one-to-one correspondence between the monomials in $\mathbb{K}[X]$ and the elements of $\enn^n$. The degree of the monomial $X^{\alpha}$ with $\alpha=(\alpha_1,\ldots,\alpha_n)$ is $\mid \alpha\mid:=\alpha_1+\alpha_2+\cdots+\alpha_n$.  Any order $>$ we establish on $\enn^n$ will give us an order on the set of monomials in $\mathbb{K}[X]$ : if $\alpha>\beta$ according to this order, we have $X^\alpha>X^\beta$.
  An order $>$ in $\mathbb{K}[X]$ is compatible with multiplication if for all $X^\alpha$, $X^\beta$ and $X^\gamma$ in $\mathbb{K}[X] $ with  $X^\alpha > X^\beta $ then $X^\alpha X^\gamma > X^\beta X^\gamma$. Now let $>$ be an order on the set of monomials  $X^u$ where $u\in\enn^n$. We say that $>$ is a  semigroup order in $\mathbb{K}[X]$ if $>$ is a total order and it is compatible with the multiplication of monomials. A monomial order on $\mathbb{K}[X]$ is a semigroup order such that $1 < X_i $ for $i= 1,\dots, n$. 
   Usual monomial orders on $\mathbb{K}[X]$ are the lexicographic order, the degree lexicographic order and the degree reverse lexicographic order.\\
   Let $f$ be a non-zero polynomial of $\mathbb{K}[X]$ such that  $f = \displaystyle{\sum_{i=1}^{k}c_{i}X^{\alpha_i}}$ where $c_{i}\in \mathbb{K}$ and  $\alpha_i\in\enn^n$. Let us fix a monomial order $>$ on $\mathbb{K}[X]$. A term in $\mathbb{K}[X]$ is a scalar times a monomial.   The leading term of $f$, denoted by $\lt_>(f)$, is the largest involved term with respect to $>$. If $\lt_>(f)= c_{m}X^{\alpha_m}$ where $1\leqslant m\leqslant k$, then $c_m$ is called the leading coefficient of $f$ ($\lc_>(f)$) and $X^{\alpha_m}$ is the leading monomial of $f$ ($\lm_>(f)$). We denote by $\deg(f)$ the maximal degree of all monomials occuring in $f$. The reduction of a polynomial $f$ by a polynomial $g$, denoted by $\red(f,g)$ is defined by \[\red(f,g) := f-q.g\] where $\lt_>(f)=q.\lt_>(g)$, for some term $q=cX^\alpha$. Let $\mathcal{F} = (f_1,...,f_s)$ be a s-tuple of polynomials in $ \mathbb{K}[X]$. Each polynomial $f\in\mathbb{K}[X]$ can be written in the form :
    \begin{equation*}
    f=a_1f_1+a_2f_2+\dots+a_sf_s+r
    \end{equation*}
    \noindent where $a_1,...,a_s$ , $r \in\mathbb{K}[X]$ and either $r=0$ or $r$ is a $\mathbb{K}$-linear combination of monomials, none of which  is divisible by any of $\lt_>(f_1),...,\lt_>(f_s)$. Moreover, if  $a_if_i\not = 0$, then $\lt_>(f)\geqslant \lt_>(a_if_i) $, $1\leqslant i\leqslant s$. The polynomial $r$ is called the remainder of $f$ on division by $\mathcal{F}$. The remainder $r$ is produced by the following algorithm called division algorithm in $\mathbb{K}[X]$ (see \cite{clo1}).
    
    \hspace{2cm}  Input :  \ \ $f_1,\dots,\  f_s,\  f$\ 
      
    \hspace{2cm}Output : \ \ $r$ \ \ \ 
    
     \hspace{2cm} $r:= 0$ \ \ 
      
      \hspace{2cm} $p:=f$ 
       
     \hspace{2cm} WHILE \ \ $p\neq 0$ \ DO \  
      
    \hspace{4cm} $i:=1$
     
       \hspace{4cm} divisionoccurred:= false
     
    \hspace{4cm} WHILE $i\leqslant s$ \ AND \ divisionoccurred = false \ DO 
    
    \hspace{5.5cm} IF $\lt_>(f_i)$\ divides\ $\lt_>(p)$\ THEN 
    
     \hspace{7cm} $p:= \red(p\  , f_i)$
     
    \hspace{7cm}  divisionoccurred:= true
     
      \hspace{5.5cm} ELSE 
      
       \hspace{7cm} $i:=i+1$
       
        \hspace{4cm} IF \ divisionoccurred = false \ THEN 
    
     \hspace{5.5cm} $r:= r+\lt_>(p)$

    \hspace{5.5cm} $p:= p-\lt_>(p)$
    
    \noindent The division algorithm terminates after a finite number of steps.\\
   Let $I\subset\mathbb{K}[X]$ be a non-zero ideal and $>$ a monomial order. The ideal generated by the set of $\lt_>(f)$ where $f\in I$ is called the  leading ideal of $I$, denoted by $\lt_>(I)$, i.e 
    \begin{equation*}
       \lt_>(I)  :=\langle   \lt_>(f)/f\in I\rangle.   
    \end{equation*}
   For a finite subset  $G=\{g_1,...,g_s\}$ of the ideal  $I$, we denote by $\lt_>(G)$ the ideal generated by the $\lt_>(g_i)$, for $i = 1,\ldots,s$, i.e
    \begin{equation*}
       \lt_>(G) :=\langle   \lt_>(g_1),\ldots,\lt_>(g_s)\rangle.   
    \end{equation*}
      A finite subset $G =\{g_1,...,g_s\}$ of an ideal $I$  is called a Groebner basis of $I$ if $\lt_>(G) =   \lt_>(I)$. A Groebner basis $G$ for the polynomial ideal $I$ is called a minimal Groebner basis of $I$  if for all $g \in G$ , $\lc_>(g)=1$ and $\lt_>(g)\not\in \lt_>(G-\{g\})$. The reduced Groebner basis for the ideal $I$ is a Groebner basis $G$ satisfying :
      
       $(1)$ $\lc_>(g)=1 $ for all $g\in G$,
    
    $(2)$ no monomial of $g$ lies in  
    $\lt_>(G-\{g\})$, for all $g\in G$.\\
    A Groebner basis can be determined by using Buchberger's algorithm. Let $f,g\in \mathbb{K}[X]$ be non-zero polynomials. Fix a monomial order $>$ and let $X^\gamma$ be the least common multiple of the leading monomial of $f$  and the leading monomial of $g$. The S-polynomial of $f$ and $g$, denoted by  $\spoly(f,g)$ is 
  \begin{center}
 $ \spoly(f,g) := \dfrac{X^\gamma}{\lt_>(f)}.f-\dfrac{X^\gamma}{\lt_>(g)}.g$
  \end{center}
  If $I$ is a polynomial ideal, then a basis 
  $G= \{g_1,\dots,g_s\}$ for $I$ is a Groebner basis for $I$ if and only if for all pairs $i\neq j$, the remainder on division of $\spoly(g_i,g_j)$ by $G$ is zero.

\section{Localization}
Let $\mathbb{K}$ be a commutative field and $p=(p_1, \dots, p_n)$  an n-tuple of  $\mathbb{K}^n$.   $\mathbb{K}(X):=\mathbb{K}(X_1, \dots, X_n)$ denotes the field of rationnal fonctions containing $\mathbb{K}[X]$. We define a local ring in $\mathbb{K}(X)$ by the set
$\mathcal{O}_p := \biggl\lbrace\dfrac{f}{g} / f, g\in \mathbb{K}[X], g(p)\not = 0\biggr\rbrace$. We say also that we localize at the maximal ideal $\langle X_1-p_1,\dots,X_n-p_n\rangle\subset\mathbb{K}[X]$.  
 Let $m_p$ be the ideal generated by $X_1-p_1,\dots,X_n-p_n$ in $\mathcal{O}_p$. Then each element in $\mathcal{O}_p\setminus m_p$ is a unit in $\mathcal{O}_p$.\\
  A local order  in $\mathbb{K}[X]=\mathbb{K}[X_1,\dots,X_n]$ is a semigroup order such that $1> X_i$ for all $1\leqslant i\leqslant n$. For instance,
for two n-tuples of $\enn^n$ $\alpha = (\alpha_1,\ldots,\alpha_n)$ and $\beta = (\beta_1,\ldots,\beta_n)$, we define the negative degree lexicographic order  by  $\alpha>\beta$ if $\mid\alpha\mid<\mid\beta\mid$ or $\mid\alpha\mid=\mid\beta\mid$ and there exists an integer $i\in\{1,\dots,n\}$ such that $\alpha_1=\beta_1,\dots,\alpha_{i-1}=\beta_{i-1},\alpha_i>\beta_i$.\\
 Let $>$ be a local order on the set of monomials of  $\mathbb{K}[X]$ and let\\
  $S_>:=\{ 1+g \in\mathbb{K}[X]\slash\  g = 0\ or\  \lt_>(g)<1 \}$. $S_>$ is a multiplicative part of $\mathbb{K}[X]$ and we remark that $S_>=\mathbb{K}[X]\setminus\langle X_1,\dots, X_n\rangle$. Define the localization of $\mathbb{K}[X]$ in view of the order $>$ by the ring (\cite{clo2, tan})
  \begin{equation}\label{loc} \loc_>(\mathbb{K}[X]):=S_>^{-1}\mathbb{K}[X]=\biggl\{ \dfrac{f}{(1+g)} / f\in \mathbb{K}[X],\ 1+g\in S_>\biggr\}. 
  \end{equation}
    We have $\loc_>(\mathbb{K}[X])=\mathcal{O}_{p=0}$. Under the local order, there is a difficulty for the successive reductions, because we may have an infinite strictly decreasing sequence of terms. For example, consider the polynomials of one variable $X$, $f =X$ and $g=X-X^2$, and we divide $f$ by $g$ by using the division algorithm, so that we successively reduce by $X-X^2$
     . This gives the
     reductions:
    \begin{align*} 
        f_1 & :=\red(f, g)=X^2\\
        f_2 & :=\red(f_1,g)=X^3\\
            & \vdots \\
        f_n & :=\red(f_{n-1},g)=X^{n+1},
    \end{align*}
     and so on.\\ Mora introduced a method to solve this problem.
     The following result can be found in \cite{clo2}.
  \begin{thm} [Mora normal form algorithm]\label{thmMora}\ \ \\
   Given non-zero polynomials $f,f_1,\ldots, f_s\in \mathbb{K}[X]$ and let $>$ be a local order. There is an algorithm which gives the polynomials  $u,a_1,\ldots, a_s,h\in\mathbb{K}[X]$ such that
 \begin{equation}
 \label{corolaire}
 uf = a_1f_1 + \cdots + a_sf_s + h
 \end{equation}
 where $\lt_>(u) = 1$  $(u = 1+g$ is unit in $\loc_>(\mathbb{K}[X]))$ , $ \lt_>(f)\geqslant\lt_>(a_i)\lt_>(f_i)$ for all $i$ with $a_i \not = 0$, and $h = 0$ or $\lt_>(h)$ is not divisible by any of $\lt_>(f_i)$.  We denote $ NF(f\mid G):=h$ with $G=\{f_1,\dots,f_s\}$, and we say that $h$ is the weak normal form of $f$ on division by $G$.
 \end{thm}
 For $f\in\mathbb{K}[X]$, we define $\ecart(f):=\deg(f)-\deg(\lt_>(f))$. 
 \noindent The remainder $h$ in \eqref{corolaire} is produced by the following algorithm called Mora's division algorithm\\
 
\hspace{2cm}$h:= f$;\ \ \ \  $L:= \{f_1,\dots, f_s\}$;\ \  $M:=\{g\in L\ :\ \lt_>(g)\mid \lt_>(h)\}$

\hspace{2cm}  WHILE \ \ $(h\neq 0\ AND \ M\neq \emptyset )$\ DO 
  
\hspace{2cm}SELECT \ \ $g\in M$ \ \ \text{with}\ \  $\ecart(g)$ \ \ \text{minimal} 

 \hspace{2cm} IF $\ecart(g)>\ecart(h)$ \ \ THEN 
  
  \hspace{2cm} \hspace{1cm} $L:= L\cup\{h\}$ 
   
 \hspace{2cm} $h:= \red(h,g)$ 
  
\hspace{2cm} IF\ $h\neq 0\ \  THEN$ 
 
   \hspace{3cm} $M:=\{g\in L\ :\ \lt_>(g)\mid \lt_>(h)\}$
 
\section{Linear codes and binomial ideals}
Let $\mathbb{F}_p$ be the finite field with $p$ elements where $p$ is a prime number. A linear code $\mathcal{C}$ of length $n$ and dimension $k$  over $\mathbb{F}_p$ is the image of a linear (injective) mapping
 \[ \psi:\mathbb{F}_p^k  \longrightarrow\mathbb{F}_p^n\]
where $ k\leqslant    n$. The elements  $x=(x_1,...,x_n)\in\mathcal{C}$ are called the codewords. The weight of a word  $x=(x_1,...,x_n)\in\mathbb{F}_p^n$  is defined by $ w_t(x) := \card\{i / x_i\not =0,1\leqslant     i\leqslant     n\}$. The minimum distance of the linear code  $\mathcal{C}$ is 
 $d := \min\{d(x,y)/ x,y\in\mathcal{C},x\not = y\}$ or   $d := \min\{ w_t(x) \slash x\in\mathcal{C},x\not = 0\}$ where $d(x,y) := \card(\{i / x_i\neq y_i \})$. We define the support of  an element $x\in \mathcal{C}$ by $\supp(x) := \{i / x_i\neq 0 \}$. A linear code $\mathcal{C}$ of length $n$ and dimension $k$ is called an $[n,k]-code$. Moreover, if the minimum distance is $d$, we say that $\mathcal{C}$ is an $[n,k,d]-code$.\\
 Let $\mathcal{C}$ be an $[n,k]-code$, $e_i=(\zeta_{i1},...,\zeta_{ik})$ where $i=1,...,k$ the canonical basis of  $\mathbb{F}_p^k$ and\\ $\psi(e_i)=(g_{i1},...,g_{in})$ . The generating matrix of $\mathcal{C}$ is the matrix of dimension  $k \times n$ defined by
 $G=(g_{ij})$ where $g_{ij}\in\mathbb{F}_p$. The linear code  $\mathcal{C}$ is represented as follows
 $\mathcal{C}=\{xG / \; \; x\in\mathbb{F}_p^k\}$.  We will say that $G$ is in standard form
  if $G=(I_k\mid M)$ where $I_k$ is the  $k \times k$ identity matrix.\\
 Let $\mathcal{C}$ be an $[n,k]-code$ over $\mathbb{F}_p$. Define the ideal associated with $\mathcal{C}$ as (\cite{borg4,ziperfect,zi14,zi2})
 \begin{equation}
 \label{ideal}
 I_{\mathcal{C}} :=\langle X^c-X^{c^\prime} \ \ \mid c-c^\prime\in\mathcal{C} \rangle + \langle X_i^p-1 \mid 1\leqslant i \leqslant n\rangle  
 \end{equation}
 where each word $c\in\mathbb{F}_p^n$ is considered as an integral vector in the monomial $X^c$. 
  Let $\mathcal{C}$ be an $[n,k]-code$ over  $\mathbb{F}_p$ and 
  \begin{equation}\label{matricestandard}
  G = (g_{ij}) = (I_k\mid M)
  \end{equation}
    a generating matrix in standard form .
  Let $m_i$ be the vector of length $n$ over $\mathbb{F}_p$ defined by
  \begin{equation}\label{mi}
   m_i = (0,\ldots, 0,p-g_{i,k+1},\ldots, p-g_{i,n})
   \end{equation}
    for $1\leqslant i\leqslant k$. We have $X^{m_i} = X_{k+1}^{p-g_{i,k+1}}X_{k+2}^{p-g_{i,k+2}}  \dots X_{n}^{p-g_{i,n}}  = \displaystyle{\prod_{j\in \supp (m_i)}^{}X_j^{p-g_{i,j}}}$. In particular, if $\supp (m_i)=\varnothing$, then $X^{m_i}=1$.
 
  \begin{thm}\label{relationgrobcodage}
  Let us take the lexicographic order on  $\mathbb{K}[X]$ with \\$X_1>X_2>\cdots >X_n$. The code ideal  $I_\mathcal{C}$ has the reduced Groebner basis
  \begin{equation}
  \label{groebner}
  \mathcal{G} = \{ X_i-X^{m_i} / 1\leqslant i\leqslant k\}\cup \{ X_i^p-1 / k+1\leqslant i\leqslant n\}.
  \end{equation}
  \end{thm}
  \pr
 A proof can be found in \cite{zi1}.
\section{Standard bases}

In this section, we will describe the analogues of Groebner bases called standard bases for the ideals in local rings by Mora's division algorithm. Given any local order $>$ on monomials in $\mathbb{K}[X]$, there is a natural extension of $>$ to $\loc_>(\mathbb{K}[X])$, which we will also denote by $>$. For any $h=\dfrac{f}{1+g}\in\loc_>(\mathbb{K}[X])$ as in \eqref{loc}, we define $\lm_>(h):=\lm_>(f)$; $\lc_>(h):=\lc_>(f)$ and $\lt_>(h):=\lt_>(f)$. 
  We fix a local order $>$ on $\loc_>(\mathbb{K}[X])$ and let $I$ be an ideal in $\loc_>(\mathbb{K}[X])$. 
    A standard basis of $I$ is a subset  $\{f_1,\ldots,f_r\}$ of $I$ such that $\lt_>(I) = \langle\lt_>(f_1),\ldots,\lt_>(f_r)\rangle$ where $\lt_>(I)$ is the ideal generated by the set of $\lt_>(f)$ with $f\in I$. 
  \begin{prop}[Product criterion]\label{prodact} \cite{greuel}\ \ \\
 Let $f,g\in\mathbb{K}[X_1,\dots,X_n] $ be polynomials such that $\lcm(\lm_>(f),\lm_>(g))=\lm_>(f).\lm_>(g)$, then
 \begin{equation}\label{product}
 NF\biggl(\spoly(f,g)\mid\{f,g\}\biggr)=0
 \end{equation}
 where $NF(-\mid-)$ is defined as in theorem\eqref{thmMora}.
  \end{prop}
 \noindent We will consider the ideals of the local ring $\loc _>(\mathbb{K}[X])$ which are generated by polynomials of $\mathbb{K}[X]$. A more general result of the following theorem can be found in \cite{greuel}.
\begin{thm}[Buchberger criterion]\label{buchbergercriterion}\ \ \\ Let $I\subset \loc_>(\mathbb{K}[X]) $ be an ideal, 
$G=\{g_1,\dots,g_s\}$ a set of polynomials of $I$ and $>$ a local order. Let $NF(-\mid -)$ be the  weak normal form as in theorem\eqref{thmMora}. Then the following are equivalent:
\begin{itemize}
\item[i)]  $G$ is a standard basis of $I$.
\item[ii)] $G$ generates $I$ and $ NF\bigr(\spoly(g_i,g_j)\mid G\bigl)=0 $ for $i, j=1,\dots,s$.
\item[iii)] $G$ generates $I$ and $NF\bigl(\spoly(g_i,g_j)\mid G_{ij}\bigr)=0$ for a suitable subset $G_{ij}\subset G$ and $i, j=1,\dots,s$.
\end{itemize}
\end{thm} 
 \noindent Let $\mathcal{C}$ be an $[n,k]-code$, the point $(1,\dots,1)$ is a zero of the code ideal $I_{\mathcal{C}}$ in the affine space over $\mathbb{F}_p$. Rather than localizing at the maximal ideal $\langle X_1-1,\dots,X_n-1\rangle$, we change coordinates to translate the point to the origin. Denote $I_{\mathcal{C}}^\prime$ the corresponding ideal, and $I:=I_{\mathcal{C}}^\prime\loc_>(\mathbb{F}_p[X])$ the ideal of $\loc_>(\mathbb{F}_p[X])$ generated by $I_{\mathcal{C}}^\prime$.
 
\begin{lem}\label{iprim} Let $p$ a prime number, we have \\
$\displaystyle {I_{\mathcal{C}}^\prime =\biggl \langle (X_i+1)+(p-1)\prod_{j\in \supp(m_i)}^{}(X_j+1)^{p-g_{i,j}}} \slash 1\leqslant i\leqslant k\biggr\rangle  + \biggl\langle (X_i+1)^p+p-1 \slash k+1\leqslant i\leqslant n\biggr\rangle $ where $g_{i,j}$ is defined in \eqref{matricestandard} and $m_i$ in \eqref{mi}.
\end{lem}
\pr 
 The ideal $I_{\mathcal{C}}$ defined in \eqref{ideal} has the reduced Groebner basis \eqref{groebner} by the theorem\eqref{relationgrobcodage} with respect to the lexicographic order on $\mathbb{K}[X]$. This is an ideal basis of $I_{\mathcal{C}}$ in $\mathbb{K}[X]$. The translation is made via the ring map $X_i\longmapsto X_i+1 $. Since $\mathbb{K}[X]\subset\loc_>(\mathbb{K}[X])$, then the claim for the translated ideal follows.\\
\qed \\
Now we present our main result
\begin{thm}
Let $\mathcal{C}$  be an $[n,k]-code$ over  $ \mathbb{F}_p$ with $p$ a prime number. Under the negative degree lexicographic order  $>$ on $\mathbb{F}_p[X]$, the ideal $I=I_{\mathcal{C}}^\prime \loc_>(\mathbb{F}_p[X])$ in $\loc_>(\mathbb{F}_p[X])$ has the standard basis 
\begin{equation}
\displaystyle{S = \biggl\{X_i -  \sum_{\substack{0\leqslant t_l\leqslant p-g_{i,j_l} (1\leqslant l \leqslant \sigma_i)\\(t_1,\dots,t_{\sigma_i})\neq(0,\dots,0)}}^{}\prod_{h=1}^{\sigma_i}\dbinom{p-g_{i,j_{h}}}{t_h} X_{j_h}^{t_{h}} \slash 1\leqslant i\leqslant k\biggr\} \cup \biggl\{X_i^p \slash k+1\leqslant i\leqslant n\biggr\}}
\end{equation}
where $\sigma_i:=\card(\supp (m_i))$.
\end{thm}
\pr
We will show that  $S$ generates $ I_{\mathcal{C}}^\prime$, then we prove that $S$ is a standard basis.\\
 \underline{For $k+1\leqslant i\leqslant n$}\\
Let $X_i^p\in S$ and we will prove that $X_i^p\in I_\mathcal{C}^\prime$. For $j\in \{1,2,\dots,p-1 \}$, the number  $\binom{p}{j}$ is a multiple of $p$ because $p$ is prime. Since we work over a field of characteristic $p$, we have
\begin{align*}
X_i^p & = \sum_{j=1}^{p-1}\binom{p}{j}X_i^j+X_i^p\\
      & = 1+\sum_{j=1}^{p-1}\binom{p}{j}X_i^j+X_i^p-1\\
      &=\binom{p}{0}X_i^0+\sum_{j=1}^{p-1}\binom{p}{j}X_i^j+ \binom{p}{p}X_i^p-1\\
      & = \sum_{j=0}^p\binom{p}{j}X_i^j-1 \\
      & = \sum_{j=0}^p\binom{p}{j}X_i^j+p-1 \\ 
      & = (X_i+1)^p+p-1 \ \ \in I_\mathcal{C}^\prime.
\end{align*}
\noindent \underline{For $1\leqslant i\leqslant k$}\\
Let $\displaystyle{X_i -  \sum_{\substack{0\leqslant t_l\leqslant p-g_{i,j_l} (1\leqslant l \leqslant \sigma_i)\\(t_1,\dots,t_{\sigma_i})\neq(0,\dots,0)}}^{}\prod_{h=1}^{\sigma_i}\dbinom{p-g_{i,j_{h}}}{t_h} X_{j_h}^{t_{h}}}\in S$. Suppose that $\supp (m_i)=\{j_1, j_2,\dots, j_{\sigma_i} \}$ with $j_1 < j_2 < \dots < j_{\sigma_i}$ and denote $A:=\displaystyle{X_i -  \sum_{\substack{0\leqslant t_l\leqslant p-g_{i,j_l} (1\leqslant l \leqslant \sigma_i)\\(t_1,\dots,t_{\sigma_i})\neq(0,\dots,0)}}^{}\prod_{h=1}^{\sigma_i}\dbinom{p-g_{i,j_{h}}}{t_h} X_{j_h}^{t_{h}}}$. We have
\begin{align*}
    & A = \displaystyle{X_i +1-\biggl[     \sum_{\substack{0\leqslant t_l\leqslant p-g_{i,j_l} (1\leqslant l \leqslant \sigma_i)\\(t_1,\dots,t_{\sigma_i})\neq(0,\dots,0)}}^{}\prod_{h=1}^{\sigma_i}\dbinom{p-g_{i,j_{h}}}{t_h} X_{j_h}^{t_{h}}+1\biggr]}\\
                          & = \displaystyle{X_i +1- \sum_{\substack{0\leqslant t_l\leqslant p-g_{i,j_l} (1\leqslant l \leqslant \sigma_i)}}^{}\prod_{h=1}^{\sigma_i}\dbinom{p-g_{i,j_{h}}}{t_h} X_{j_h}^{t_{h}}}\\
   & =X_i +1- \displaystyle{\sum_{t_1=0}^{p-g_{i,j_{1}}}\sum_{t_2=0}^{p-g_{i,j_{2}}}\dots \sum_{t_{\sigma_i}=0}^{p-g_{i,j_{\sigma_i}}}}\dbinom{p-g_{i,j_{1}}}{t_1}\dbinom{p-g_{i,j_{2}}}{t_2}\dots\dbinom{p-g_{i,j_{\sigma_i}}}{t_{\sigma_i}} X_{j_1}^{t_{1}}X_{j_2}^{t_{2}}\dots X_{j_{\sigma_i}}^{t_{\sigma_i}}\\
                   & =  \displaystyle{X_i +1-\biggl[\sum_{t_1=0}^{p-g_{i,j_{1}}}\dbinom{p-g_{i,j_{1}}}{t_1}X_{j_1}^{t_{1}}\biggr]} \displaystyle{\biggl[\sum_{t_2=0}^{p-g_{i,j_{2}}}\dbinom{p-g_{i,j_{2}}}{t_2}X_{j_2}^{t_{2}}\biggr]}\dots\displaystyle{\biggl[\sum_{t_{\sigma_i}=0}^{p-g_{i,j_{\sigma_i}}}\dbinom{p-g_{i,j_{\sigma_i}}}{t_{\sigma_i}}X_{j_{\sigma_i}}^{t_{\sigma_i}}\biggr]}\\
  & =\displaystyle{X_i +1-\biggl[\big(X_{j_1}+1\big)^{p-g_{i,j_{1}}}\biggr]\biggl[\big(X_{j_2}+1\big)^{p-g_{i,j_{2}}}\biggr]\dots\biggl[\big(X_{j_{\sigma_i}}+1\big)^{p-g_{i,j_{\sigma_i}}}\biggr]}\\
                    & =  \displaystyle{X_i +1-\prod_{j\in\supp (m_i)}\big(X_{j}+1\big)^{p-g_{i,j}}}\\
& =  \displaystyle{X_i +1+p\biggl[\prod_{j\in\supp (m_i)}\big(X_{j}+1\big)^{p-g_{i,j}}\biggr]-\prod_{j\in\supp (m_i)}\big(X_{j}+1\big)^{p-g_{i,j}}}\\
                           & =  \displaystyle{X_i +1+(p-1)\biggl[\prod_{j\in \supp (m_i)}^{}(X_j+1)^{p-g_{i,j}}\biggr]}\in I_{\mathcal{C}}^\prime
\end{align*}
By lemma \eqref{iprim}, $S$ is a generating set for $I_{\mathcal{C}}^\prime$.\\
Let us now show that $S$ is a standard basis over  $\mathbb{F}_p$ 
\begin{itemize}
\item[*] Let the pair $(i,j)$ such that      $k+1\leqslant i <  j \leqslant n$. \\
We have $\spoly(X_i^p , X_j^p)=\dfrac{X_i^pX_j^p}{X_i^p}X_i^p-\dfrac{X_i^pX_j^p}{X_j^p}X_j^p = X_i^pX_j^p-X_i^pX_j^p = 0$, and $ NF(0 \mid S)=0 $.
\item[*] Let the pair  $(i,j)$ such that $1\leqslant i \leqslant k$ and  $k+1\leqslant j \leqslant n$. \\
 Denote \hspace{1cm}$f_i:=\displaystyle{X_i -  \sum_{\substack{0\leqslant t_l\leqslant p-g_{i,j_l} (1\leqslant l \leqslant \sigma_i)\\(t_1,\dots,t_{\sigma_i})\neq(0,\dots,0)}}^{}\prod_{h=1}^{\sigma_i}\dbinom{p-g_{i,j_{h}}}{t_h} X_{j_h}^{t_{h}}}$. 
We have
\begin{align*}
 \spoly(f_i , \ X_j^p) &= \dfrac{X_iX_j^p}{X_i}\biggl[\displaystyle{X_i -  \sum_{\substack{0\leqslant t_l\leqslant p-g_{i,j_l} (1\leqslant l \leqslant \sigma_i)\\(t_1,\dots,t_{\sigma_i})\neq(0,\dots,0)}}^{}\prod_{h=1}^{\sigma_i}\dbinom{p-g_{i,j_{h}}}{t_h} X_{j_h}^{t_{h}}} \biggr]-\dfrac{X_iX_j^p}{X_j^p}X_j^p\\
              & =X_iX_j^p-X_j^p\biggl[\displaystyle{\sum_{\substack{0\leqslant t_l\leqslant p-g_{i,j_l} (1\leqslant l \leqslant \sigma_i)\\(t_1,\dots,t_{\sigma_i})\neq(0,\dots,0)}}^{}\prod_{h=1}^{\sigma_i}\dbinom{p-g_{i,j_{h}}}{t_h} X_{j_h}^{t_{h}}} \biggr]-X_iX_j^p\\
 &=-X_j^p\biggl[\displaystyle{\sum_{\substack{0\leqslant t_l\leqslant p-g_{i,j_l} (1\leqslant l \leqslant \sigma_i)\\(t_1,\dots,t_{\sigma_i})\neq(0,\dots,0)}}^{}\prod_{h=1}^{\sigma_i}\dbinom{p-g_{i,j_{h}}}{t_h} X_{j_h}^{t_{h}}} \biggr]
\end{align*}
In the last expression, all these monomials are multiple of  $X_j^p\in S$. Therefore the remainder of the division of $\spoly(f_i , X_j^p)$ by $\{X_j^p\}$ is zero, i.e \\$NF\biggl(\spoly(f_i , X_j^p)\mid \{X_j^p\} \biggr)=0$.
\item[*] Finally, let  $1\leqslant i < j \leqslant k$.\\
 Let $g_i:=\displaystyle{X_i -  \sum_{\substack{0\leqslant t_l\leqslant p-g_{i,r_l} (1\leqslant l \leqslant \sigma_i)\\(t_1,\dots,t_{\sigma_i})\neq(0,\dots,0)}}^{}\prod_{h=1}^{\sigma_i}\dbinom{p-g_{i,r_{h}}}{t_h} X_{r_h}^{t_{h}}}$ \\ and $g_j:=\displaystyle{X_j -  \sum_{\substack{0\leqslant t_l^\prime\leqslant p-g_{j,s_l} (1\leqslant l \leqslant \sigma_j)\\(t_1^\prime,\dots,t_{\sigma_j}^\prime)\neq(0,\dots,0)}}^{}\prod_{u=1}^{\sigma_j}\dbinom{p-g_{j,s_{u}}}{t_u^\prime} X_{s_u}^{t_{u}^\prime}}$.\\
We have \hspace{0.5cm}$\lt_>(g_i)=X_i$ \hspace{0.5cm} and \hspace{0.5cm}$\lt_>(g_j)=X_j$, \hspace{0.4cm}\\ then $\lcm (\lm_>(g_i),\lm_>(g_j))=\lm_>(g_i).\lm_>(g_j)$. According to the Product Criterion in Proposition \eqref{prodact}, we obtain \\$ NF\biggl(\spoly(g_i,g_j)\mid\{g_i,g_j\}\biggr)=0$ . And by the Buchberger's criterion in theorem \eqref{buchbergercriterion}, the assertion follows.
\end{itemize}
\qed \\
 \begin{ex}
       Consider the generator matrix $G=(g_{ij})$ defined by
  \begin{quotation}
         \[
         G=
         \begin{pmatrix}
         1& 0 & 0 & 1 & 0 & 1\\
         0& 1 & 0 & 2 & 1 & 0\\
         0& 0 & 1 & 2 & 2 & 1\\
         \end{pmatrix}
         \]
         \end{quotation}
 Under the negative degree lexicographic order $>$ on $\mathbb{F}_3[X]$, the ideal  $I=I_{\mathcal{C}}^\prime  \loc_>(\mathbb{F}_3[X])$ where $I_{\mathcal{C}}^\prime$ is defined in lemma \eqref{iprim} with $p=3$, $n=6$ and $k=3$ has the standard basis 
 
 \[ \displaystyle{S = \biggl\{g_1,\ g_2,\ g_3,\ X_4^3,\ X_5^3,\ X_6^3 \biggr\}}\] 
 where \\
$g_1 = X_1+X_4+X_6+2X_4^2+2X_4X_6+2X_6^2+X_4^2X_6+X_4X_6^2+2X_4^2X_6^2$, 

 \vspace{0.3cm}\noindent $g_2 = X_2+2X_4+X_5+X_4X_5+2X_5^2+2X_4X_5^2$, 
 
\vspace{0.3cm}\noindent $g_3 = X_3+2X_4+2X_5+X_6+2X_4X_5+X_4X_6+X_5X_6+2X_6^2+X_4X_5X_6+2X_4X_6^2+2X_5X_6^2+2X_4X_5X_6^2$. 
                        
\end{ex}
    An immediate consequence is the result in \cite{f2} ( for $p=2$)
\begin{thm}In view of the negative degree lexicographic order  $>$ on $\mathbb{F}_2[X]$, the ideal
 $I = I_{\mathcal{C}}^\prime Loc _>(\mathbb{F}_2[X])$  in $Loc _>(\mathbb{F}_2[X])$ where $\displaystyle {I_{\mathcal{C}}^\prime = \langle (X_i+1)+\prod_{j\in \supp(m_i)}^{}(X_j+1)} / 1\leqslant i\leqslant k\rangle $ \\$ + \langle (X_i+1)^2+1 / k+1\leqslant i\leqslant n\rangle $  has the standard basis :
\begin{equation*}
\displaystyle{S = \biggl\{X_i -  \sum_{\substack{J\subseteq \supp(m_i)\\J\neq \varnothing}}^{}X_J / 1\leqslant i\leqslant k\biggr\} \cup \biggl\{X_i^2 / k+1\leqslant i\leqslant n\biggr\}}.
\end{equation*}
\end{thm}

\end{document}